\documentclass[sigconf]{acmart}

\usepackage[dvipsnames]{xcolor}
\usepackage{xspace}
\usepackage{url}
\usepackage{graphicx}
\usepackage{breakurl} 
\usepackage{cleveref}
\usepackage{enumitem}
\usepackage{multirow}
\usepackage{color}
\usepackage{colortbl}
\usepackage{balance}
\usepackage{tabularx}

\newcommand{\totalStudentsIncludingNotFinished}{34\xspace}

\newcommand{\totalStudents}{20\xspace}

\newcommand{\totalStudentsInterview}{6\xspace}

\newcounter{observation}
\newcommand{\observation}[1]{\refstepcounter{observation}
	\begin{center}
		\framebox{
			\begin{minipage}{0.93\columnwidth}
				{} \textit{#1}
			\end{minipage}
		}
	\end{center}
}

\graphicspath{ {./img/} }

\title[Understanding Student Interaction with AI-Powered Next-Step Hints: Strategies and Challenges]{Understanding Student Interaction with AI-Powered\\Next-Step Hints: Strategies and Challenges}


\AtBeginDocument{%
  }

\copyrightyear{2026}
\acmYear{2026}
\setcopyright{cc}
\setcctype{by}
\acmConference[SIGCSE TS 2026]{Proceedings of the 57th ACM Technical Symposium on Computer Science Education V.1}{February 18--21, 2026}{St. Louis, MO, USA}
\acmBooktitle{Proceedings of the 57th ACM Technical Symposium on Computer Science Education V.1 (SIGCSE TS 2026), February 18--21, 2026, St. Louis, MO, USA}
\acmPrice{}
\acmDOI{10.1145/3770762.3772544}
\acmISBN{979-8-4007-2256-1/2026/02}

\begin{document}

\author{Anastasiia Birillo} 
\affiliation{
  \institution{JetBrains Research}
  \city{Belgrade}
  \country{Serbia}
}
\email{anastasia.birillo@jetbrains.com}

\author{Aleksei Rostovskii} 
\affiliation{
  \institution{JetBrains Research}
  \city{Berlin}
  \country{Germany}
}
\email{aleksei.rostovskii@jetbrains.com}

\author{Yaroslav Golubev} 
\affiliation{
  \institution{JetBrains Research}
  \city{Belgrade}
  \country{Serbia}
}
\email{yaroslav.golubev@jetbrains.com}

\author{Hieke Keuning} 
\affiliation{
  \institution{Utrecht University}
  \city{Utrecht}
  \country{Netherlands}
}
\email{h.w.keuning@uu.nl}

\begin{abstract}

Automated feedback generation plays a crucial role in enhancing personalized learning experiences in computer science education. Among different types of feedback, next-step hint feedback is particularly important, as it provides students with actionable steps to progress towards solving programming tasks. This study investigates how students interact with an AI-driven next-step hint system in an in-IDE learning environment. We gathered and analyzed a dataset from \totalStudentsIncludingNotFinished students solving Kotlin tasks, containing detailed hint interaction logs. We applied process mining techniques and identified 16 common interaction scenarios. Semi-structured interviews with \totalStudentsInterview students revealed strategies for managing unhelpful hints, such as adapting partial hints or modifying code to generate variations of the same hint. These findings, combined with our publicly available dataset, offer valuable opportunities for future research and provide key insights into student behavior, helping improve hint design for enhanced learning support.

\end{abstract}


\maketitle

\section{Introduction}
\label{section:introduction}

Automated feedback plays a key role in personalized learning in education, including computer science. By providing timely and tailored guidance, it helps students better understand programming concepts and overcome challenges more effectively~\cite{jeuring2022towards, marwan2020adaptive, vanlehn2011relative}. Automated feedback can come in various forms~\cite{paiva2022automated}, such as code quality feedback~\cite{birillo2022hyperstyle, iddon2023gradestyle, keuning2021tutoring}, enhanced compiler error messages~\cite{leinonen2023using, phung2023generating}, next-step hint feedback~\cite{birillo2024one, xiao2024exploring, rivers2013automatic, liffiton2023codehelp}, and more.

Next-step feedback is particularly important for helping students progress with coding tasks. The goal of this feedback is to provide a single actionable \textit{step} that brings the student's solution closer to passing all tests for a given task~\cite{keuning2018systematic}. Over the years, the methods for generating next-step hints have evolved significantly, moving from traditional data-driven~\cite{rivers2013automatic, its2017isnap} and rule-based approaches~\cite{its2017haskell} to utilizing large language models (LLMs)~\cite{roest2024next, liu2024teaching, aruleba2023integrating} combined with other techniques like static analysis~\cite{birillo2024one}. 

Recently, the \textit{in-IDE} learning format was introduced, with the primary goal of placing students in a professional environment, teaching not only programming concepts but also the use of professional tools~\cite{birillo2024bridging}. From a technical perspective, this format offers many research opportunities, as it leverages powerful IDE features integrated with existing educational methods. For instance, our recent paper~\cite{birillo2024one} proposed combining static analysis and code quality checks available in the IDE with LLMs to deliver next-step hints while controlling their size and quality. 
While the system showed promising results through expert validation and classroom testing~\cite{birillo2024one}, the study primarily evaluated tool quality rather than student interactions. In addition to studying how successful or unsuccessful these interactions really are, it is of crucial importance to learn how the students deal with the less useful hints. 
In this study, we therefore aim to answer two research questions:

\begin{enumerate}[leftmargin=1.22cm,start=1,label={\bfseries RQ\arabic*:}]
    \item What behavioral patterns can be identified based on students' interaction with the next-step hint system?
    \item What strategies do students employ to overcome unhelpful hints?
\end{enumerate}

\vspace{-2mm}

To explore these questions, we collected fine-grained data on the problem-solving processes of \totalStudentsIncludingNotFinished first- and second-year Bachelor’s students completing Kotlin programming tasks, capturing keystrokes, IDE actions, accessed windows, and detailed hint interactions. The full dataset includes 6,658,936 lines of code, 1,364,943 IDE activity events, 960 textual hint requests, and 453 code hint requests, and it is publicly available for future research~\cite{supplementary}. Based on this data, we performed a quantitative analysis of student behavior. We used the \textit{process mining} technique and identified 16 common interaction scenarios.
To gain deeper insights, we conducted 30-minute interviews with \totalStudentsInterview students who completed all tasks. We uncovered strategies students used for overcoming unhelpful hints, such as selecting parts of hints to adapt them or modifying their code to generate multiple variations of the same hint, which they integrated into their solutions.

\section{Related Work \& Background}
\label{section:related_work}

\subsection{Prior Research on Hint Usage}

Many studies have explored the use of feedback to support students working on programming tasks~\cite{jeuring2022towards, keuning2018systematic}. Recently, several works have specifically focused on how students interact with hint systems~\cite{xiao2024exploring, bui2024hints, wiggins2021exploring}.

Wiggins et al.~\cite{wiggins2021exploring} examined an intelligent next-step hint system, focusing on \textit{why} students seek help. They studied the elapsed time and code completeness after each used hint, and identified reasons for seeking hints such as struggling to start an exercise or improving existing code. While the study outlines general hint-seeking trends, it does not explore how hints influence problem-solving strategies.

Jeuring et al.~\cite{jeuring2022towards} studied \textit{when} and \textit{how} to provide feedback and hints during programming tasks, using datasets of students' keystroke data. The data was annotated to show where experts would intervene and how feedback would be given. However, the study focuses on expert-driven interventions and does not explore how students themselves interact with hint systems.

Xiao et al.~\cite{xiao2024exploring} analyzed how various hint levels aid students in solving tasks, and also identified reasons \textit{why} students seek hints. Through a think-aloud study with pre- and post-tests involving 12 students, they examined the role of hints in learning. They found that high-level text hints alone can be unhelpful or even misleading, and adding low-level hints, such as code hints, might better support students. However, the study does not explore broader hint usage patterns or strategies for handling unhelpful hints.

In a recent study, Bui et al.~\cite{bui2024hints} examined how students interact with on-demand automated hints and their impact on learning in an SQL course. Using A/B testing, they analyzed differences in student behavior and identified simple patterns, such as most students requesting a second hint within 10 seconds of the first. While providing useful insights into basic behavioral trends, the primary focus of the study was to determine whether the inclusion of a hint system effectively supports student learning, rather than analyzing the identified common interaction patterns in detail.

While prior studies examined \textit{why} and \textit{when} hints are provided, little research focuses on \textit{how} students interact with hints. This can be achieved with the \textit{process mining} method, which transforms log data into behavioral models through action analysis~\cite{bannert2014process, bogarin2018survey}. Though common in other educational fields~\cite{ghazal2017educational, boroujeni2018discovery}, this method was not yet applied to programming hint systems. This study applies process mining to uncover behavioral patterns in student-hint interactions.

\subsection{JetBrains Academy \& Its AI-Based Hints}
\label{section:context}

This subsection outlines the context of the current study, focusing on the JetBrains Academy plugin designed for in-IDE learning and the AI-based next-step hint system integrated within the plugin.

\textbf{JetBrains Academy Plugin.} The JetBrains Academy plugin, designed for JetBrains IDEs, was developed to support the new concept of \textit{in-IDE learning}~\cite{birillo2024bridging}. This format of learning proposes integrating the learning process (\textit{i.e.}, theoretical lessons and quizzes, as well as practical programming tasks) directly into the professional IDE environment. 
A detailed description of the plugin functionality can be found in our previous work~\cite{birillo2024bridging}. 

\textbf{AI-Based Next-Step Hint System.} In another recent paper, we introduced a novel approach to generating next-step hints for the JetBrains Academy plugin~\cite{birillo2024one}.  
It provides two levels of next-step hints: \textit{textual} and \textit{code} hints.
When a student faces issues with a coding task, such as not knowing what to do next, a failed test, or a compilation error, they can request help by pressing the \textit{Get Hint} button. 
Upon doing so, a textual hint appears, outlining one actionable step towards a solution, with relevant lines highlighted directly in the student's code. 
If this hint is insufficient for the student, they can request a code hint, which shows the exact code fragment required to bring their solution closer to the correct one.

The novelty of this approach is a combination of state-of-the-art LLMs with static analysis techniques. Static analysis is utilized to control the size of the generated code hints, to make sure they are granular enough. To improve the overall hint quality, the textual hint is generated from the code hint (and showed to the student first, as described above). Consequently, the static analysis modifications applied to the code hint are also used in generating the textual hint, ensuring they remain of suitable size. While the system showed promising results through expert validation and classroom testing~\cite{birillo2024one}, specific student interactions were not studied. In this work, we address this gap by collecting and analyzing log data, as well as conducting and analyzing interviews about student interactions with the hint system.

\section{Data Collection}
\label{section:data:collection}

In our study, we used two types of data: (1) detailed logs of the task-solving process collected with the help of the KOALA tool~\cite{karol2025koala} and (2) interview transcripts derived from semi-structured interviews.

\subsection{Data Collection Setup}

\textbf{Programming Tasks}. For this study, we used the programming tasks from the \textit{Kotlin Onboarding: Introduction} course~\cite{kotlin-onboarding-introduction}, which are the same as the original study that introduced the hint system~\cite{birillo2024one}.
The course consists of six console-based projects designed to teach basic programming concepts in Kotlin. Participants were required to complete three out of six projects, selected to represent varying levels of complexity. The projects cover basic topics such as variables, loops, conditional operators, null-safety, and functions. We expected that the completion of all three projects should take the participants up to six hours in total. Participants were allowed to use the hint system as frequently as needed while solving the tasks.

\textbf{Participants}. We invited students to participate by emailing those who took part in our previous studies and presenting this study during several programming courses at universities. The study involved first- and second-year computer science Bachelor's students who had previously completed at least one or two programming courses in any programming language other than Kotlin, and had no prior experience with Kotlin. As a reward, students received Amazon gift cards. The logs were collected from \totalStudentsIncludingNotFinished students as they worked on solving three programming projects, with \totalStudents out of them completing all three required projects and the rest completing only part of the required tasks. Additionally, 30-minute semi-structured interviews were conducted with \totalStudentsInterview out of \totalStudents participants who completed all the tasks. These interviewees were selected to represent a varied range of programming experience levels and different locations to ensure more diverse and comprehensive insights. 

\textbf{Data Privacy \& Ethics}. Participation in this study was voluntary. Students could choose to take part only in the task-solving process or also in the interview. They were informed about the study’s purpose, data collection, and data usage. According to the consent form signed by the students, personal data (\textit{e.g.}, names and emails) will be deleted two years after the study, and interview recordings will be removed, keeping only anonymized transcripts for analysis.

\subsection{Task Solving Logs}

The task-solving data was collected using the KOALA tool~\cite{karol2025koala}. The tool was configured to capture all keystroke-level code changes within task files, interactions within the IDE (\textit{e.g.}, program executions or test runs), and all interactions with the hint system. In total, we collected 6,658,936 code lines, 1,364,943 IDE activity events, 960 textual hint requests, and 453 code hint requests, which were used for the analysis. The collected data is available as one of the contributions of this work~\cite{supplementary}.

\subsection{Interview Data}

The first author conducted \totalStudentsInterview semi-structured interviews, each lasting approximately 30 minutes. The interviews started with an introduction and ended with a summary, with the main section focusing on participants' experiences with the AI-based next-step hints. The primary goal of the interviews was to gain a deeper understanding of students' experiences with the hint system. The full interview script is available in the supplementary materials~\cite{supplementary}. 

\begin{table*}[t]
\centering
\caption{The scenarios of hint requests in our logs. All scenarios start with asking for a hint (\textcolor{Cerulean}{HBC}) or retrying a hint (\textcolor{Cerulean}{HRC}), positive ones end with the hint being accepted (\textcolor{OliveGreen}{HA}), negative ones end with canceling the hint (\textcolor{Mahogany}{HCAN}) or an error (\textcolor{Mahogany}{ERR}).}
\label{tab:action_patterns_updated}
\begin{tabular}{@{}lll@{}}
\toprule
\multicolumn{1}{c}{\textbf{Scenario}}      & \multicolumn{1}{c}{\textbf{Sequence of Actions}}                       & \multicolumn{1}{c}{\textbf{Frequency}} \\ \midrule
\multicolumn{3}{c}{\textcolor{OliveGreen}{\textbf{POSITIVE Scenarios}}} \\ 
\midrule
Scenario Pos-1       & \textcolor{Cerulean}{HBC} $\rightarrow$ HCG $\rightarrow$ HBS $\rightarrow$ SCHC $\rightarrow$ \textcolor{OliveGreen}{HA}     &  293 (27.90\%)  \\
Scenario Pos-2       & \textcolor{Cerulean}{HBC} $\rightarrow$ HCG $\rightarrow$ HBS $\rightarrow$ SCHC $\rightarrow$ HBCLO $\rightarrow$ \textcolor{OliveGreen}{HA}     & 2 \ \ \ \ (0.19\%) \\
Scenario Pos-3       & \textcolor{Cerulean}{HRC} $\rightarrow$ HCG $\rightarrow$ HBS $\rightarrow$ SCHC $\rightarrow$ \textcolor{OliveGreen}{HA} & 4 \ \ \ \ (0.38\%)  \\
\rowcolor{gray!20}\textbf{Total positive}             &                                                               & \textbf{299 (28.48\%)} \\ \midrule
 
\multicolumn{3}{c}{\textcolor{YellowOrange}{\textbf{NEUTRAL Scenarios}}} \\\midrule
Scenario Neut-1        & \textcolor{Cerulean}{HBC} $\rightarrow$ HCG $\rightarrow$ HBS                         & 263 (25.05\%)  \\
Scenario Neut-2        & \textcolor{Cerulean}{HRC} $\rightarrow$ HCG $\rightarrow$ HBS     & 2 \ \ \ \ (0.19\%)  \\
Scenario Neut-3        & \textcolor{Cerulean}{HBC} $\rightarrow$ HCG $\rightarrow$ HBS $\rightarrow$ SCHC & 88 \ \ (8.38\%)  \\
Scenario Neut-4        & \textcolor{Cerulean}{HRC} $\rightarrow$ HCG $\rightarrow$ HBS $\rightarrow$ SCHC     & 5 \ \ \ \ (0.48\%)   \\
Scenario Neut-5        & \textcolor{Cerulean}{HBC} $\rightarrow$ HCG $\rightarrow$ HBS $\rightarrow$ SCHC $\rightarrow$ HBCLO & 18 \ \ (1.71\%)   \\
Scenario Neut-6        & \textcolor{Cerulean}{HBC} $\rightarrow$ HCG $\rightarrow$ HBS $\rightarrow$ HBCLO  & 42 \ \ (4.00\%)   \\
Scenario Neut-7        & \textcolor{Cerulean}{HRC} $\rightarrow$ HCG $\rightarrow$ HBS $\rightarrow$ HBCLO  & 2 \ \ \ \ (0.19\%)  \\
\rowcolor{gray!20}\textbf{Total neutral}             &                                                              & \textbf{420 (40.00\%)}  \\ \midrule

\multicolumn{3}{c}{\textcolor{Mahogany}{\textbf{NEGATIVE Scenarios}}} \\\midrule
Scenario Neg-Cancel-1    & \textcolor{Cerulean}{HBC} $\rightarrow$ HCG $\rightarrow$ HBS $\rightarrow$ SCHC $\rightarrow$ \textcolor{Mahogany}{HCAN}      & 9 \ \ \ \ (0.86\%) \\
Scenario Neg-Cancel-2    & \textcolor{Cerulean}{HBC} $\rightarrow$ HCG $\rightarrow$ HBS $\rightarrow$ SCHC $\rightarrow$ HBCLO $\rightarrow$ \textcolor{Mahogany}{HCAN} & 9 \ \ \ \ (0.86\%) \\ 
Scenario Neg-Err-1     & \textcolor{Cerulean}{HBC} $\rightarrow$ HBS $\rightarrow$ \textcolor{Mahogany}{ERR}                        & 167 (15.90\%) \\
Scenario Neg-Err-2     & \textcolor{Cerulean}{HBC} $\rightarrow$ HBS $\rightarrow$ \textcolor{Mahogany}{ERR} $\rightarrow$ HBCLO    & 31 \ \ (2.95\%)  \\
Scenario Neg-Err-3     & \textcolor{Cerulean}{HRC} $\rightarrow$ HBS $\rightarrow$ \textcolor{Mahogany}{ERR}    & 55 \ \ (5.24\%)  \\
Scenario Neg-Err-4     & \textcolor{Cerulean}{HRC} $\rightarrow$ HBS $\rightarrow$ \textcolor{Mahogany}{ERR} $\rightarrow$ HBCLO    & 13 \ \ (1.24\%)   \\
\rowcolor{gray!20}\textbf{Total negative}             &                                                              & \textbf{284 (27.05\%)} \\ \midrule

\multicolumn{3}{c}{\textcolor{Gray}{\textbf{OTHER Scenarios}}} \\\midrule
Other scenarios                     &                                                              & 47 \ \ (4.48\%)  \\ \bottomrule
\end{tabular}
\end{table*}

\section{Analysis \& Results}
\label{section:analysis}

\subsection{RQ1: Student Behavior Patterns}

In RQ1, we analyzed the logs to establish student behavioral patterns when using the hint system.

\textbf{Scenarios to Request Hints}. We preprocessed the collected log data by extracting all \textit{hint-request sessions} from the dataset. A \textit{hint-request session} is a sequence of actions performed by a student related to a single hint. Each session begins when the student presses the \textit{Get Hint} or \textit{Retry} buttons (the \textit{Retry} button generates a new hint for the same code and task). A session ends when there is evidence that the hint was processed by the student, such as accepting a code hint, closing a text hint banner, or encountering an error. The session also concludes if no further activity occurred within the next five minutes.

Second, we divided all hint-request sessions into four possible ways to interact with the system: \textit{positive}, \textit{neutral}, \textit{negative}, and \textit{other}. \textit{Positive} refer to cases where students explicitly accepted hints by clicking the \textit{Accept} button after receiving the code hint. \textit{Neutral} occur when the text hint was displayed to students, but the students did not clearly indicate their intention to accept or reject it. These ways of interactions are investigated in a more detailed way in RQ2 (Section~\ref{sec:rq3}). \textit{Negative} take place when students either declined hints by clicking the \textit{Cancel} button or encountered system errors during hint generation. Finally, there are a few \textit{other} ways to interact that require further investigation, as we could not determine the type of session based on the available data. 

To define the scenarios within each way of interaction, we considered the following actions as part of interacting with the hint system.
\begin{itemize}
    \item \textbf{HBC (Hint Button Clicked)}: The student clicks the \textit{Get Hint} button to generate a hint.
    \item \textbf{HRC (Hint Retry Clicked)}: The student clicks the \textit{Retry} button to regenerate a hint.
    \item \textbf{HCG (Hint Code Generated)}: The code hint is generated. By tool design, this action is triggered every time the HBC action is performed.
    \item \textbf{HBS (Hint Banner Shown)}: The banner with a textual hint or a system's error message is shown to the student. Although this action is carried out by the system, not by the student, it marks the moment when the student views the banner. This action will be used for a more detailed analysis in RQ2.
    \item \textbf{SCHC (Show Code Hint Clicked)}: The student clicks the \textit{Show Code} button to show a code hint.
    \item \textbf{HA (Hint Accepted)}: The student clicks the \textit{Accept} button to accept a code hint.
    \item \textbf{HBCLO (Hint Banner Closed)}: The student closes the textual hint banner.
    \item \textbf{HCAN (Hint Canceled)}: The student clicks the \textit{Cancel} button to decline a code hint.
    \item \textbf{ERR (Error Occurred)}: This action includes cases where internal errors occurred during the hint generation process, \textit{e.g.}, problems with internet connection, or the LLM not providing any response.
\end{itemize}

With this list of actions, we defined scenarios based on the possible ways students could interact with the hint system, where each scenario starts with requesting or retrieving a hint. The final list of scenarios was agreed between the first two authors of the paper. Then, we ran a Python script to extract all of them from the collected data. A detailed description of all scenarios can be found in~\Cref{tab:action_patterns_updated}. We observed that the total number of positive and neutral scenarios is 719, accounting for 68.48\% of all scenarios. Although neutral scenarios do not \textit{directly} indicate the acceptance of the hint, we assume that in most cases they are likely positive, since the students observed the textual hint and did not explicitly cancel the code one. Specifically, \textit{Scenario Neut-1}, \textit{Scenario Neut-2}, \textit{Scenario Neut-6}, and \textit{Scenario Neut-7} indicate that students did not request a code hint at all, as receiving only the textual hint most likely was sufficient for them to progress with the task. 

Most of the negative scenarios (266 out of 284) occurred due to internal system errors. We analyzed and categorized the error messages to identify the most frequent problems. Our investigation revealed that only two main issues occurred: (1) Internet connection problems on the student’s side, and (2) internal system errors related to interactions with LLM providers and access tokens. Both of these issues can be resolved within the hint system and do not have a significant impact on the overall quality of the hints.

\begin{figure*}[t]
    \centering
    \includegraphics[width=0.8\linewidth]{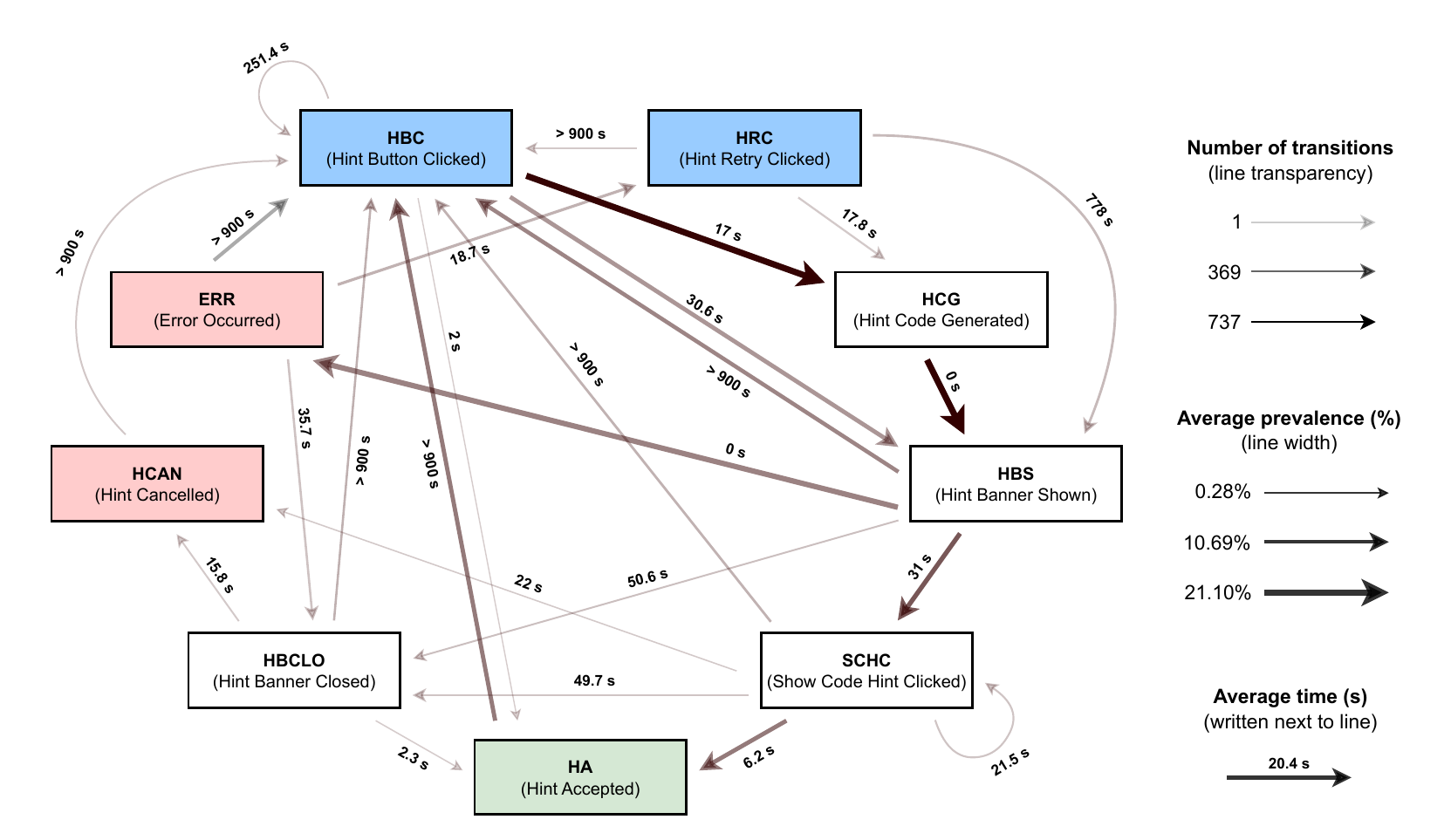}
     \vspace{-0.3cm}
    \caption{A process behavior graph showing hint request scenarios across all students. Large times (\textit{e.g.}, {\textgreater}900 seconds) indicate that the student was solving the task and then returned for another hint (\textit{i.e.}, a new scenario started).}
     \vspace{-0.3cm}
    \label{fig:process_graph:all}
\end{figure*}

\textbf{Process mining}. \textit{Process mining} translates complex log data into interpretable behavioral models, considering both the frequency and order of actions~\cite{bannert2014process, bogarin2018survey}. After identifying hint request scenarios, we visualized them as \textit{a process behavior graph}. 
First, we added the actions from the list above as nodes in the graph.
Then, we included all transitions between the graph nodes, including self-transitions (loops).
Specifically, we calculated: (1) the number of times each transition occurred in total; (2) the average prevalence of the transition, \textit{i.e.}, the percentage that this transition takes up from all studied transitions, averaged between all students; and (3) the average transition time in seconds. 
The resulting graph can be seen in~\Cref{fig:process_graph:all}. For transitions longer than 15 minutes, we used the label \textbf{{\textgreater}900s} to improve clarity. It can be seen that all such transitions lead towards \textit{HBC}, thus indicating cases where students switched between scenarios from \Cref{tab:action_patterns_updated}. More specifically, this shows that students stopped using the system for some time and then returned to request another hint.

The  graph shows several interesting findings. First, we can see that transition \textit{ERR} $\rightarrow$ \textit{HBC} is common for students, which means that students continued using the system even after encountering an error. However, they did not proceed with requesting a new hint immediately after the error, as the average time between these two actions was more than 15 minutes. Instead, they returned to the system later. Interestingly, this behavior of continuing to use the system after encountering an error was relatively common among students, accounting for an average of 9.8\% of all transitions.
These findings are important, as these numbers show that students continued to use the system even when they encountered difficulties. This is significant in real-world scenarios, where systems often have bugs and issues, especially when LLMs play a major role.

We also observed that some students struggled with using the code hint feature, as they clicked the \textit{Show code hint} button repeatedly within a short time frame (\textit{SCHC} $\rightarrow$ \textit{SCHC},  average time 21.49 seconds). This might suggest that students were unable to locate the code hint window or were unsure how to use it. However, this issue requires further investigation to better understand the cause.

\observation{\textbf{Summary of RQ1.} Among all the studied interactions with the system, 28.48\% are positive, 40\% are neutral, and 27.05\% are negative. Most negative interactions occur because of system errors that can be overcome in the future. After errors, students usually wait for some time until requesting a new hint later in the course, or quickly try to regenerate the hint. Students also sometimes press buttons several times, seemingly not understanding them.}

\subsection{RQ2: Student Strategies for Unhelpful Hints}
\label{sec:rq3}

It can be seen from RQ1 that there are quite a few neutral and negative scenarios for students, and yet they come back for more hints. Because of this, we decided to ask participants how they handled unhelpful hints and what specific strategies they used.

\textbf{Preparation of Interview Data}. We used Google Meet to conduct and record the interviews. The tool automatically generated transcripts of the recordings, which the first author reviewed and corrected for any inaccuracies. 

\textbf{Results of Interview Data Analysis}. The first author conducted an initial analysis of the responses about handling unhelpful hints. This analysis aimed to identify possible strategies students used to deal with unhelpful hints and to determine which scenarios from RQ1 required further, detailed investigation.

We found that two out of six interviewees chose not to use the \textit{Accept} button for the code hints and instead worked with the code hints manually. In some cases, this was done to better remember what they were doing, while in others, they wanted to apply only part of the hint --- the \textit{Selective Use of Hints} strategy. For this case we will focus on scenarios, where students requested a code hint but chose not to apply it directly.

An interesting insight from two other students is combining information from multiple hints to troubleshoot and develop a working solution. They made small changes to their solutions to generate new hints and repeated this process several times without accepting these hints --- the \textit{Combining Partial Solutions} strategy. This approach is not common for the system and requires further analysis. To identify these scenarios, we need to examine sequential neutral scenarios within a single task that occur within a short time interval. To learn more about these cases from both strategies, we returned to the collected log data from the specific interviewees who reported them, and studied them in detail.

First, let's explore the \textbf{Selective Use of Hints} strategy, where students manually applied only a part of the given code hints to their solutions. We focused on \textit{Scenario Neut-3} (occurring a total of 10 times for these students), where students requested a hint, asked for a code hint, and did not continue further.
The first author of the paper manually reviewed the logs for these 10 occurrences, including textual and code hints, student actions after receiving a hint, code changes, and IDE interactions. 

\begin{figure}[t]
    \centering
    \includegraphics[width=0.65\linewidth]{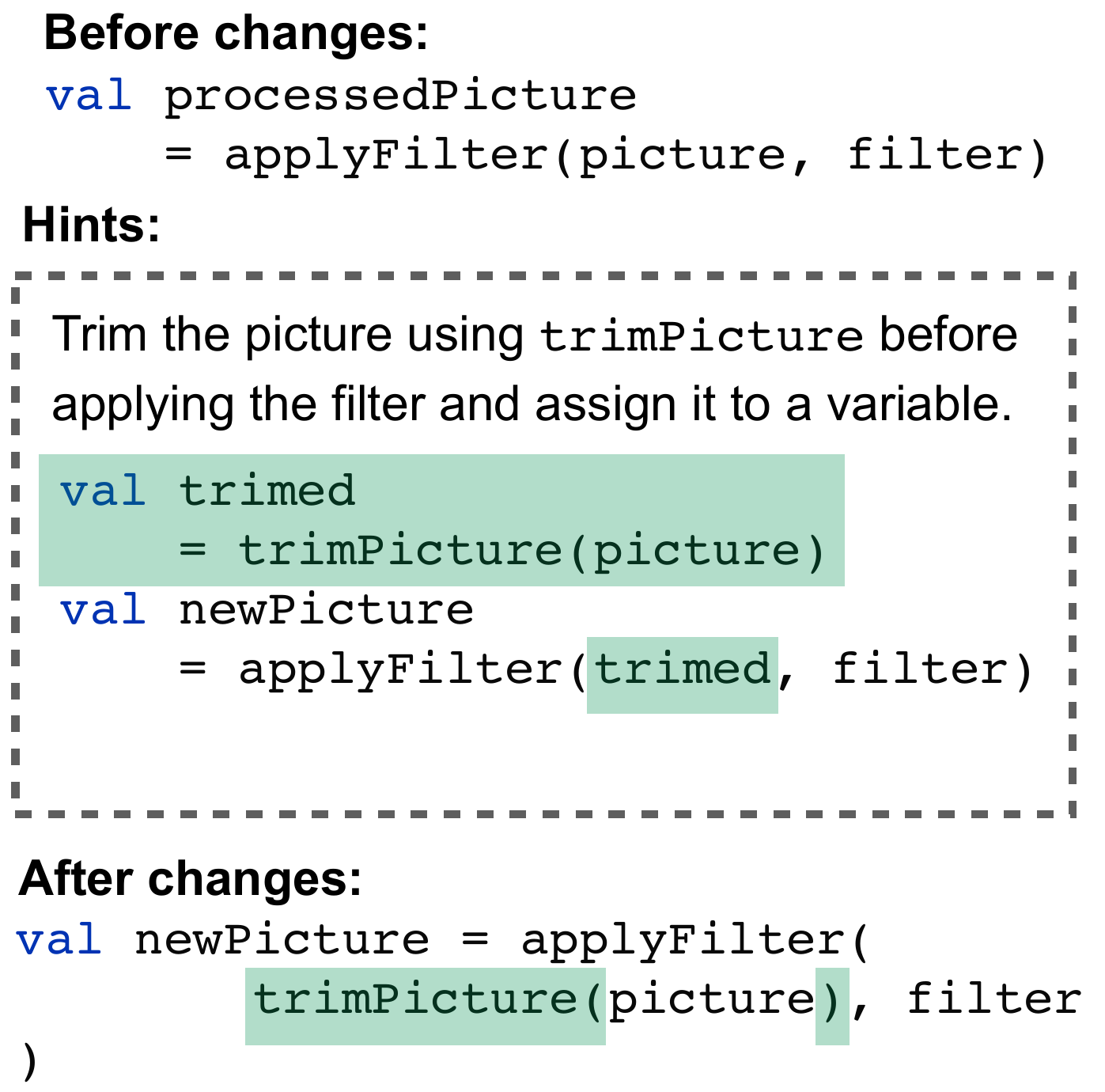}
     \vspace{-0.3cm}
    \caption{An example of selective use of hints strategy: a student adapts the suggestion and applies it partially in their code. ``Before changes'' shows the student's initial code before requesting a hint. ``Hints'' shows both the text and code hints for the student's code. ``After changes'' shows the student's code after adapting and applying the hints.}
    \label{fig:selective_use:trim}
\end{figure}

We found that in most cases after viewing the text hint, students opened the diff window in the code editor, where they could view both their own file and the code hint, and then retyped the recommended code from the hint. However, when the hint was unclear or not optimal, students demonstrated different behaviors. For example, in one case (see \Cref{fig:selective_use:trim}), a text hint recommended trimming an image before applying filters by storing a temporary result in a variable. The student followed only the first part of this suggestion, trimming the image without creating a variable. \textit{This shows that students in our study do not blindly follow the code hint instructions but instead analyze them and adapt them to get an even better solution.}

\begin{figure}[t]
    \centering
    \includegraphics[width=0.7\linewidth]{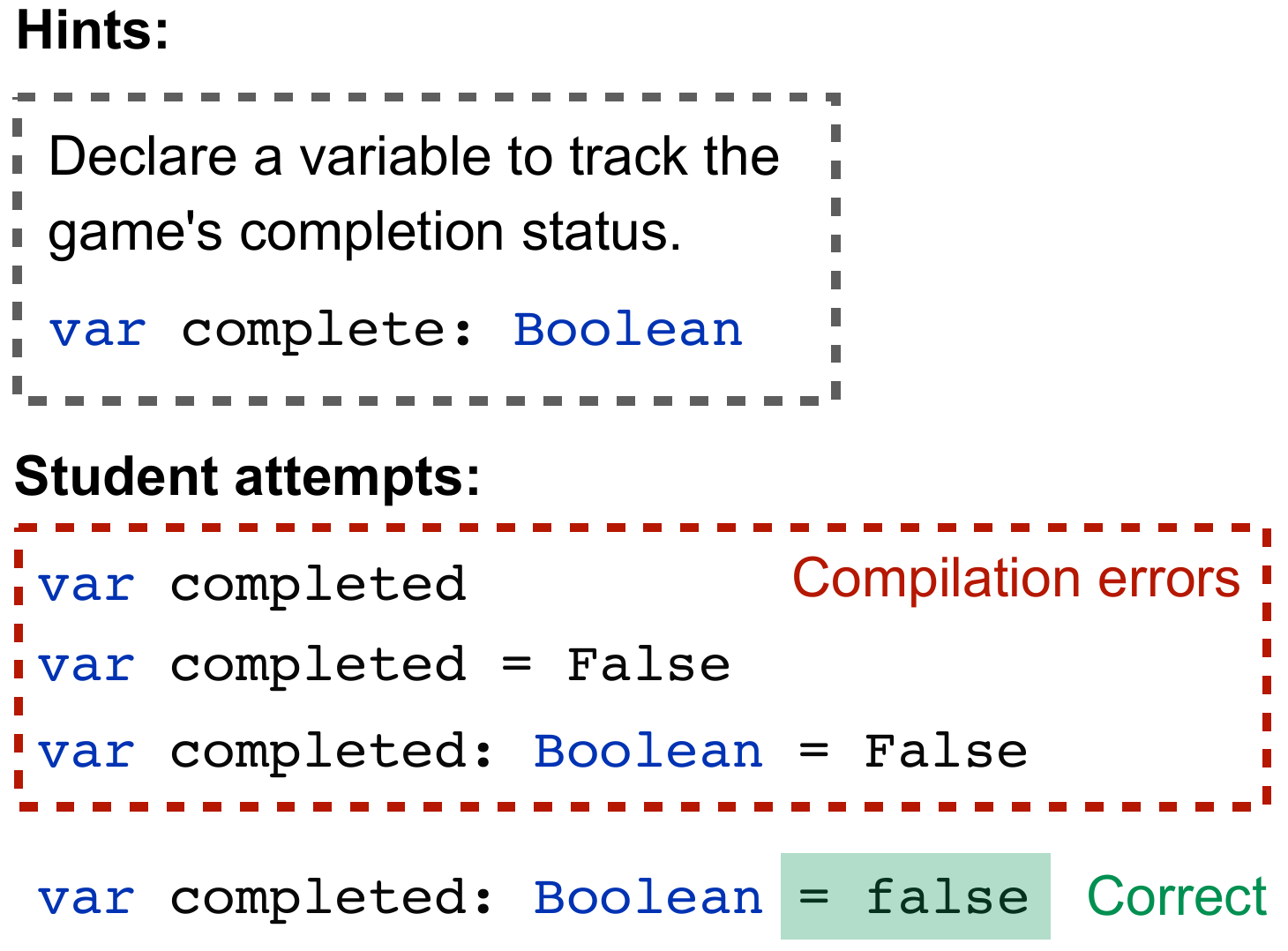}
     \vspace{-0.3cm}
    \caption{An example of selective use of hints strategy: a student tries different combinations to find a correct one. ``Hints'' shows both the text and code hints for the student's code. ``Student attempts'' shows several attempts to apply the hints. The section below shows the final (correct) attempt by the student to apply the hints.}
    \label{fig:selective_use:errors}
\end{figure}

In another case (see \Cref{fig:selective_use:errors}), a hint instructed the student to add a boolean variable to store the result of the game. In Kotlin, if such a variable is used in a loop, it might be initialized later, so the hint provided the code without an initial value. The student copied only the part with defining a variable without a type but then struggled to initialize it, making several attempts to find the working solution. \textit{This indicates that students likely need more explanations about language-specific features in the hint.}

Now let us turn our attention to \textbf{Combining Partial Solutions}, where students made small adjustments to their code to generate several hints, which they then combined to progress with the task. 

The first author conducted three steps of manual analysis on the hint logs from the students interviewed: (1) extracting the task, hint, and exact time and date; (2) grouping the extracted hints by task and date, and sorting them by time; (3) identifying sequential similar text hint requests that relate to the same piece of code. We identified 10 sequences for further analysis, consisting of two to six consecutive hint requests.

Next, the first author checked the following conditions: (1) the first hint did not directly solve the problem, i.e., was \textit{unhelpful}; (2)~the student applied their own changes based on the recommended code rather than using the recommended code directly. To validate this, the first author manually checked each group of hints. We found that 3 out of the 10 groups satisfied all the conditions. Two of these three groups came from the same task, which was the most difficult one.

Finally, for each of these three groups of hints, we analyzed the students' code changes throughout the entire hint request process. We observed that students often returned to the version of their code for which they initially requested a hint, and then requested a new hint. This was likely because they found the hint not useful enough and wanted to explore another possible variation. Additionally, students combined insights from multiple hints while refining their code. This was possible because they \textit{kept previous code hint windows open for reference}. This suggests that students may use the systems in unexpected ways, and it is important to investigate this further.

According to the interview data, students often copied their attempts outside the task environment and reused earlier attempts alongside new ones, ultimately getting a working solution in all the cases we analyzed. This behavior suggests another possible direction for future research on interaction with such systems: \textit{providing students with multiple hints with different approaches simultaneously, which could help them construct a working solution more effectively.}

\observation{\textbf{Summary of RQ2.} In neutral scenarios, students show unusual ways of interacting with the hint system. In some cases, the neutral scenario provided a helpful hint, but students manually applied only a part of the proposed solution to make it even better. In other cases, students tried to overcome unhelpful hints by requesting the hint several times, observing different suggestions, and combining them.}
\section{Threats to Validity}
\label{section:threats}

The main limitation of our study is the small number of participants, both for data collection and interviews. However, we made an effort to include students with diverse backgrounds from different countries. Despite the small sample size, we were able to gather over one million IDE events and nearly one thousand hint requests. We believe this large and detailed dataset allows for in-depth qualitative analysis with a smaller number of participants.

Additionally, this study was conducted in an in-IDE environment using the Kotlin programming language. However, the hint usage scenarios can be generalized to similar systems, as actions like \textit{hint button clicked} or \textit{hint banner shown} are not specific to any IDE or programming language. Student strategies, such as \textit{selective use of hints}, are also likely relevant to other languages.

\section{Conclusion}
\label{section:conclusion}

In this study, we analyzed how students interacted with an AI-powered hint system integrated into an in-IDE learning environment for programming tasks. By examining the scenarios of students interacting with the system and interviewing them, we aimed to understand how they engaged with different types of hints --- textual and code hints --- and how they responded to cases when the hints are unhelpful. Our analysis involved fine-grained data collection from \totalStudentsIncludingNotFinished students, process mining of detailed task-solving sequences, and collecting insights from interviews with \totalStudentsInterview students who completed all tasks.

The findings offer valuable insights into student behavior with next-step hints and highlight opportunities for enhancing such systems to better support learning. We also contribute a publicly available dataset on AI-powered hint usage to support this goal.

\bibliographystyle{ACM-Reference-Format}
\balance
\bibliography{ref}

\end{document}